\documentclass[aps,prb]{revtex4}
\usepackage{amssymb}
\usepackage{graphicx}
\usepackage{dcolumn}
\usepackage{bm}

\begin{document}

\title{Josephson effect and quantum merging of two Bose superfluids}
\author{Hongwei Xiong$^{1,2,3,\thanks{%
Electronic address:~xionghongwei@wipm.ac.cn}}$, Shujuan Liu$^{1,2,3}$,
Mingsheng Zhan$^{1,2} $}

\address{$^{1}$State Key Laboratory of Magnetic Resonance
and Atomic and Molecular Physics, Wuhan Institute of Physics and
Mathematics, Chinese Academy of Sciences, Wuhan 430071, P. R.
China}
\address{$^{2}$Center for Cold Atom Physics, Chinese Academy of Sciences, Wuhan 430071, P.
R. China}
\address{$^{3}$Graduate School of the Chinese Academy of
Sciences, P. R. China}
\date{\today }

\begin{abstract}
We consider the Josephson effect when two independent Bose superfluids are
weakly connected. In the presence of interparticle interaction and based on
the calculations of the one-particle density matrix of the whole system, we
find that the one-particle density matrix can be factorized which satisfies
the general criterion of Bose superfluid proposed by Penrose and Onsager. By
introducing an effective order parameter for the whole system, our
researches show that there is Josephson effect for two independent Bose
superfluids.

\end{abstract}

\maketitle

\section{\protect\bigskip introduction}

In 1962, Brian Josephson predicted the Josephson effect \cite{Josephson} by
considering the quantum effect of two superconductors separated by a thin
insulator. The Josephson effect is a natural quantum phenomenon for two
coherently superposed macroscopic quantum objects. The subsequent
experiments verified the Josephson effect and also lent support to the BCS
theory about the physical mechanism of superconductors. The Josephson effect
has been used in the invention of novel devices for extremely
high-sensitivity measurements of currents, voltages, and magnetic fields. Up
to date, there are still considerable interests in the fundamental physics
and applications of this effect. Similarly to the case of superconductors,
the ideal Josephson effect was also observed by investigating the flow of
superfluid $^{4}He$ through an array of micro-apertures \cite{Helium}.
Recent experiments \cite{Albiez,Anker} also observed clear Josephson current
and especially self-trapping phenomenon \cite{Smerzi} for Bose condensates
in dilute gases.

About twenty years ago, P. W. Anderson raised a famous question \cite%
{Anderson} that when two initially separated superfluids are connected,
whether the two superfluids would show a relative phase and therefore
Josephson current. In the ordinary physical picture of Josephson effect, the
initial quantum state is regarded as a coherent superposition of two
macroscopic phase coherent quantum objects. In Anderson's question, however,
before connecting two superfluids, the two superfluids are completely
independent. To answer Anderson's question, we shall consider the problem
that during the connection process or with the development of time after the
connection process, whether the quantum state of the whole system will
become a coherent superposition of two macroscopic quantum superfluids.

After the experimental realization of Bose-Einstein condensate in dilute
gases \cite{BEC}, there are a lot of interesting studies about the splitting
of a condensate and merging of two independent condensates, which relate
closely to the above Anderson's question. The former question about the
formation of a fragmented condensate during the splitting process of a
condensate has been studied intensively \cite{Spekk,Capuzzi,Javan,Menotti}.
The quantum merging of two independent condensates (i.e., the inverse
process of the splitting) was also investigated recently by considering
carefully the role of dissipation \cite{Zapata}, and by considering both the
adiabatic and nonadiabatic merging \cite{Yi} based on the well-known two
mode approximation \cite{Milburn}. Most recently, the splitting and the
following merging of an elongated condensate \cite{Mebrahtu} is considered
by including the finite-temperature effect. On the experimental side, a
continuous condensate source was created by periodically replenishing a
condensate with new condensates \cite{Chikk}. This striking experiment also
gives strong motivation to study theoretically the merging of independent
condensates.

To describe the essential quantum feature of a Bose superfluid, Penrose and
Onsager \cite{Penrose} proposed the idea of off-diagonal long-range order
(ODLRO) which gives the general criterion of a Bose superfluid. The ODLRO
plays a very important role in the description of the Bose superfluid,
especially because it has no classical analog \cite{Penrose,Yang}. For a
single superfluid, if the one-particle density matrix $\langle \widehat{\Psi
}^{\dagger }\left( \mathbf{r}_{1},t\right) \widehat{\Psi }\left( \mathbf{r}%
_{2},t\right) \rangle $ can be factorized, \textit{i.e.}, $\langle \widehat{%
\Psi }^{\dagger }\left( \mathbf{r}_{1},t\right) \widehat{\Psi }\left(
\mathbf{r}_{2},t\right) \rangle =\Phi ^{\ast }\left( \mathbf{r}_{1},t\right)
\Phi \left( \mathbf{r}_{2},t\right) $, there is an ODLRO for the superfluid,
and the superfluid can be regarded as a macroscopic quantum object which has
stable spatial coherence property \cite{Penrose}. For two initially
independent superfluids, here we investigate theoretically the dynamic
process of the whole system when the barrier separating two superfluids is
decreased adiabatically so that the two initially separated and independent
superfluids are weakly connected. Based on the calculations of the
one-particle density matrix of the whole system, it is found that there is
an interaction-induced quantum merging process for initially independent
superfluids. After two initially independent superfluids merge into a single
condensate, it is shown that $\langle \widehat{\Psi }^{\dagger }\left(
\mathbf{r}_{1},t\right) \widehat{\Psi }\left( \mathbf{r}_{2},t\right)
\rangle \simeq \Phi _{e}^{\ast }\left( \mathbf{r}_{1},t\right) \Phi
_{e}\left( \mathbf{r}_{2},t\right) $ with $\Phi _{e}$ being an effective
order parameter of the whole system. The effective order parameter has the
property that it is a coherent superposition of two macroscopic wave
functions. For weakly interacting Bose superfluids, we find that the
evolution of $\Phi _{e}$ can be described very well by the ordinary
Gross-Pitaevskii (GP) equation \cite{GP}, and thus there is also Josephson
current when two initially independent superfluids are weakly connected.

The paper is organized as follows. In Section II we introduce the effective
order parameter of the whole system when the one-particle density matrix is
calculated for the general situation. In Section III we give the expression
of the overall energy and the evolution equations based on the action
principle. A brief summary and discussion is given in Section IV.

\section{\protect\bigskip effective order parameter for two initially
independent Bose superfluids}

For two independent Bose superfluids at zero temperature shown in Fig. 1a,
the number of particles $N_{1}$ and $N_{2}$ in each of the two superfluids
are fixed, and the corresponding quantum state is
\begin{equation}
\left\vert N_{1},N_{2}\right\rangle =\frac{C_{n}}{\sqrt{N_{1}!N_{2}!}}(%
\widehat{a}_{1}^{\dag })^{N_{1}}(\widehat{a}_{2}^{\dag })^{N_{2}}\left\vert
0\right\rangle ,  \label{initial-state}
\end{equation}%
where $C_{n}$ is a normalization constant to assure $\left\langle
N_{1},N_{2}|N_{1},N_{2}\right\rangle =1$. $\widehat{a}_{1}^{\dag }$ ($%
\widehat{a}_{2}^{\dag }$) is a creation operator which creates a particle
described by the single-particle state $\phi _{1}$ ($\phi _{2}$) in the left
(right) superfluid. In this initial quantum state, the quantum depletion is
omitted. Thus, this initial quantum state is valid when $a/\overline{l}<<1$
with $a$ and $\overline{l}$ being respectively the scattering length and
mean distance between particles. One should note that for two initially
coherently-separated superfluids, the state is $\left\vert N\right\rangle =(%
\widehat{b}^{\dagger })^{N}\left\vert 0\right\rangle /\sqrt{N!}$ with $%
\widehat{b}^{\dagger }$ being a creation operator which creates a particle
with the single-particle state $\left( \sqrt{N_{1}}\phi _{1}+\sqrt{N_{2}}%
\phi _{2}\right) /\sqrt{N}$ with $N=N_{1}+N_{2}$. As shown in Fig. 1b, we
consider the dynamic evolution process of the whole system when two
initially independent superfluids are weakly connected\textit{.}

\begin{figure}[tbp]
\includegraphics[width=0.5\linewidth,angle=270]{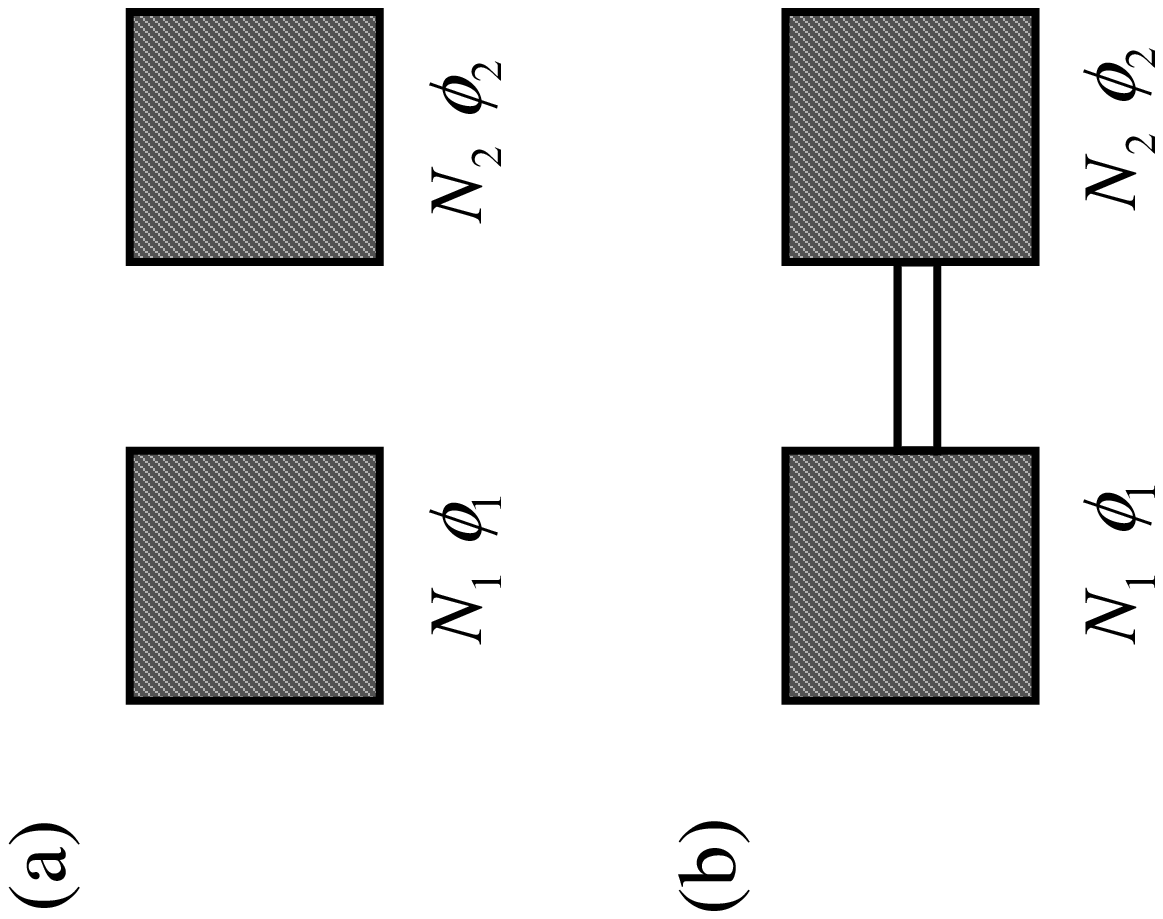} \center
\caption{Shown in Fig. 1a is two independent Bose superfluids in two
separate tanks. After the two Bose superfluids are connected shown in Fig.
1b, there is particle current between two tanks. }
\end{figure}

As shown in the following, after two independent Bose superfluids are
connected, $\phi _{1}$ and $\phi _{2}$ will overlap and especially become
non-orthogonal in the presence of interparticle interaction. Thus, we
consider the general case for $\int \phi _{1}\phi _{2}^{\ast }dV=\zeta $
from the beginning. We first give the general expression of the one-particle
density matrix $\langle \widehat{\Psi }^{\dagger }\left( \mathbf{x},t\right)
\widehat{\Psi }\left( \mathbf{y},t\right) \rangle $ for this general case.
Generally speaking, $\left\vert \zeta \right\vert <<1$. However, after
straightforward calculations, it is shown clearly that the nonzero value of $%
\zeta $ can play important role in the one-particle density matrix for $%
N_{1}\left\vert \zeta \right\vert >1$ and $N_{2}\left\vert \zeta \right\vert
>1$.

The operators $\widehat{a}_{1}$ and $\widehat{a}_{2}$ can be written as $%
\widehat{a}_{1}=\int \widehat{\Psi }\phi _{1}^{\ast }dV$ and $\widehat{a}%
_{2}=\int \widehat{\Psi }\phi _{2}^{\ast }dV$, respectively. Here $\widehat{%
\Psi }\left( \mathbf{x},t\right) $ is the field operator. By using the
commutation relations of the field operators $[\widehat{\Psi }\left( \mathbf{%
x},t\right) ,\widehat{\Psi }\left( \mathbf{y},t\right) ]=0$ and $[\widehat{%
\Psi }\left( \mathbf{x},t\right) ,\widehat{\Psi }^{\dagger }\left( \mathbf{y}%
,t\right) ]=\delta \left( \mathbf{x}-\mathbf{y}\right) $, it is easy to get
the commutation relation $[\widehat{a}_{1},\widehat{a}_{2}^{\dagger }]=\zeta
^{\ast }$. We see that $\widehat{a}_{1}$ and $\widehat{a}_{2}^{\dagger }$
are not commutative any more for $\int \phi _{1}\phi _{2}^{\ast }dV$ being a
nonzero value.

It is well-known that the field operator should be expanded in terms of a
complete and orthogonal basis set. Generally speaking, the field operator $%
\widehat{\Psi }\left( \mathbf{x},t\right) $ can be expanded as:
\begin{equation}
\widehat{\Psi }\left( \mathbf{x},t\right) =\widehat{a}_{1}\phi _{1}\left(
\mathbf{x},t\right) +\widehat{k}\phi _{2}^{\prime }\left( \mathbf{x}%
,t\right) +\cdots ,  \label{new-expansion}
\end{equation}%
where $\phi _{1}$ and $\phi _{2}^{\prime }$ are orthogonal normalization
wave functions. Assuming that $\phi _{2}^{\prime }=\beta \left( \phi
_{2}+\alpha \phi _{1}\right) $, based on the conditions $\int \phi
_{1}^{\ast }\phi _{2}^{\prime }dV=0$ and $\int \left\vert \phi _{2}^{\prime
}\right\vert ^{2}dV=1$, we have $\left\vert \beta \right\vert =\left(
1-\left\vert \zeta \right\vert ^{2}\right) ^{-1/2}$ and $\alpha =-$ $\zeta
^{\ast }$. Based on $\widehat{k}=\int \widehat{\Psi }\left( \phi
_{2}^{\prime }\right) ^{\ast }dV$, we have $\widehat{a}_{2}=\widehat{k}%
/\beta ^{\ast }+\zeta \widehat{a}_{1}$.

Because $\widehat{k}$ and $\widehat{a}_{1}^{\dagger }$ are commutative, it
is convenient to calculate the one-particle density matrix $\rho \left(
\mathbf{x},\mathbf{y},t\right) $ using the operators $\widehat{k}$ and $%
\widehat{a}_{1}^{\dagger }$. The exact expression of $\rho \left( \mathbf{x},%
\mathbf{y},t\right) $ is
\begin{eqnarray}
\rho \left( \mathbf{x},\mathbf{y},t\right) &=&\left\langle
N_{1},N_{2},t\right\vert \widehat{\Psi }^{\dag }\left( \mathbf{x},t\right)
\widehat{\Psi }\left( \mathbf{y},t\right) \left\vert
N_{1},N_{2},t\right\rangle  \nonumber \\
&=&\gamma _{1}\phi _{1}^{\ast }\left( \mathbf{x},t\right) \phi _{1}\left(
\mathbf{y},t\right) +\gamma _{2}e^{i\varphi _{c}}\phi _{1}^{\ast }\left(
\mathbf{x},t\right) \phi _{2}\left( \mathbf{y},t\right)  \nonumber \\
&&+\gamma _{2}e^{-i\varphi _{c}}\phi _{2}^{\ast }\left( \mathbf{x},t\right)
\phi _{1}\left( \mathbf{y},t\right) +\gamma _{3}\phi _{2}^{\ast }\left(
\mathbf{x},t\right) \phi _{2}\left( \mathbf{y},t\right) ,
\label{fin-density-matrix}
\end{eqnarray}%
where the coefficients are
\begin{eqnarray}
\gamma _{1} &=&\sum\limits_{i=0}^{N_{2}}\frac{C_{n}^{2}N_{2}!\left(
N_{1}+i-1\right) !N_{1}\left( 1-\left\vert \zeta \right\vert ^{2}\right)
^{N_{2}-i}\left\vert \zeta \right\vert ^{2i}}{i!i!\left( N_{1}-1\right)
!\left( N_{2}-i\right) !},  \nonumber \\
\gamma _{2} &=&\sum\limits_{i=0}^{N_{2}-1}\frac{C_{n}^{2}N_{2}!\left(
N_{1}+i\right) !\left( 1-\left\vert \zeta \right\vert ^{2}\right)
^{N_{2}-i-1}\left\vert \zeta \right\vert ^{2i+1}}{i!\left( i+1\right)
!\left( N_{1}-1\right) !\left( N_{2}-i-1\right) !},  \nonumber \\
\gamma _{3} &=&\sum\limits_{i=0}^{N_{2}-1}\frac{C_{n}^{2}N_{2}!\left(
N_{1}+i\right) !\left( 1-\left\vert \zeta \right\vert ^{2}\right)
^{N_{2}-i-1}\left\vert \zeta \right\vert ^{2i}}{i!i!N_{1}!\left(
N_{2}-i-1\right) !}.
\end{eqnarray}%
In addition, the normalization constant is determined by%
\begin{equation}
C_{n}^{2}\left( \sum\limits_{i=0}^{N_{2}}\frac{N_{2}!\left( N_{1}+i\right)
!\left( 1-\left\vert \zeta \right\vert ^{2}\right) ^{N_{2}-i}\left\vert
\zeta \right\vert ^{2i}}{i!i!N_{1}!\left( N_{2}-i\right) !}\right) =1.
\label{normconst}
\end{equation}%
The phase factor $\varphi _{c}$ is determined by $\zeta =\left\vert \zeta
\right\vert e^{i\varphi _{c}}$.

The above one-particle density matrix is obtained based on the second
quantization method. We have also proven that this one-particle density
matrix is equal to the result calculated from the many-body wave function
which satisfies the exchange symmetry of identical bosons.

Introducing the ordinary order parameter $\Phi _{1}\left( \mathbf{x}%
,t\right) =\sqrt{N_{1}}\phi _{1}\left( \mathbf{x},t\right) $ and $\Phi
_{2}\left( \mathbf{x},t\right) =\sqrt{N_{2}}\phi _{2}\left( \mathbf{x}%
,t\right) $, the one-particle density matrix can be naturally rewritten as
\begin{equation}
\rho \left( \mathbf{x},\mathbf{y},t\right) =\rho _{fac}\left( \mathbf{x},%
\mathbf{y},t\right) +\rho _{non}\left( \mathbf{x},\mathbf{y},t\right) .
\label{fact-nonf}
\end{equation}%
In the above equation, the factorable component $\rho _{fac}\left( \mathbf{x}%
,\mathbf{y},t\right) $ is
\begin{equation}
\rho _{fac}\left( \mathbf{x},\mathbf{y},t\right) =\Phi _{e}^{\ast }\left(
\mathbf{x},t\right) \Phi _{e}\left( \mathbf{y},t\right) ,  \label{factor-com}
\end{equation}%
where we have introduced an effective order parameter
\begin{equation}
\Phi _{e}\left( \mathbf{x},t\right) =\sqrt{\gamma _{1}^{\prime }}\Phi
_{1}\left( \mathbf{x},t\right) +\sqrt{\gamma _{3}^{\prime }}e^{i\varphi
_{c}}\Phi _{2}\left( \mathbf{x},t\right) .  \label{effective-order}
\end{equation}%
The coefficients are $\gamma _{1}^{\prime }=\gamma _{1}/N_{1}$, $\gamma
_{2}^{\prime }=\gamma _{2}/\sqrt{N_{1}N_{2}}$ and $\gamma _{3}^{\prime
}=\gamma _{3}/N_{2}$. The non-factorable component $\rho _{non}\left(
\mathbf{x},\mathbf{y},t\right) $ is
\begin{equation}
\rho _{non}\left( \mathbf{x},\mathbf{y},t\right) =\eta \left( e^{i\varphi
_{c}}\Phi _{1}^{\ast }\left( \mathbf{x},t\right) \Phi _{2}\left( \mathbf{y}%
,t\right) +e^{-i\varphi _{c}}\Phi _{2}^{\ast }\left( \mathbf{x},t\right)
\Phi _{1}\left( \mathbf{y},t\right) \right) ,  \label{non-cond}
\end{equation}%
where $\eta =\left( \gamma _{2}^{\prime }-\sqrt{\gamma _{1}^{\prime }\gamma
_{3}^{\prime }}\right) $. We see that the parameter $\eta $ shows the
proportion of the non-factorable component. If the coefficient $\eta $ is
approximate to $0$, $\rho _{non}\left( \mathbf{x},\mathbf{y},t\right) $ can
be omitted, and thus $\rho \left( \mathbf{x},\mathbf{y},t\right) $ can be
approximated as the factorable component $\rho _{fac}\left( \mathbf{x},%
\mathbf{y},t\right) $.

\begin{figure}[tbp]
\includegraphics[width=0.6\linewidth,angle=270]{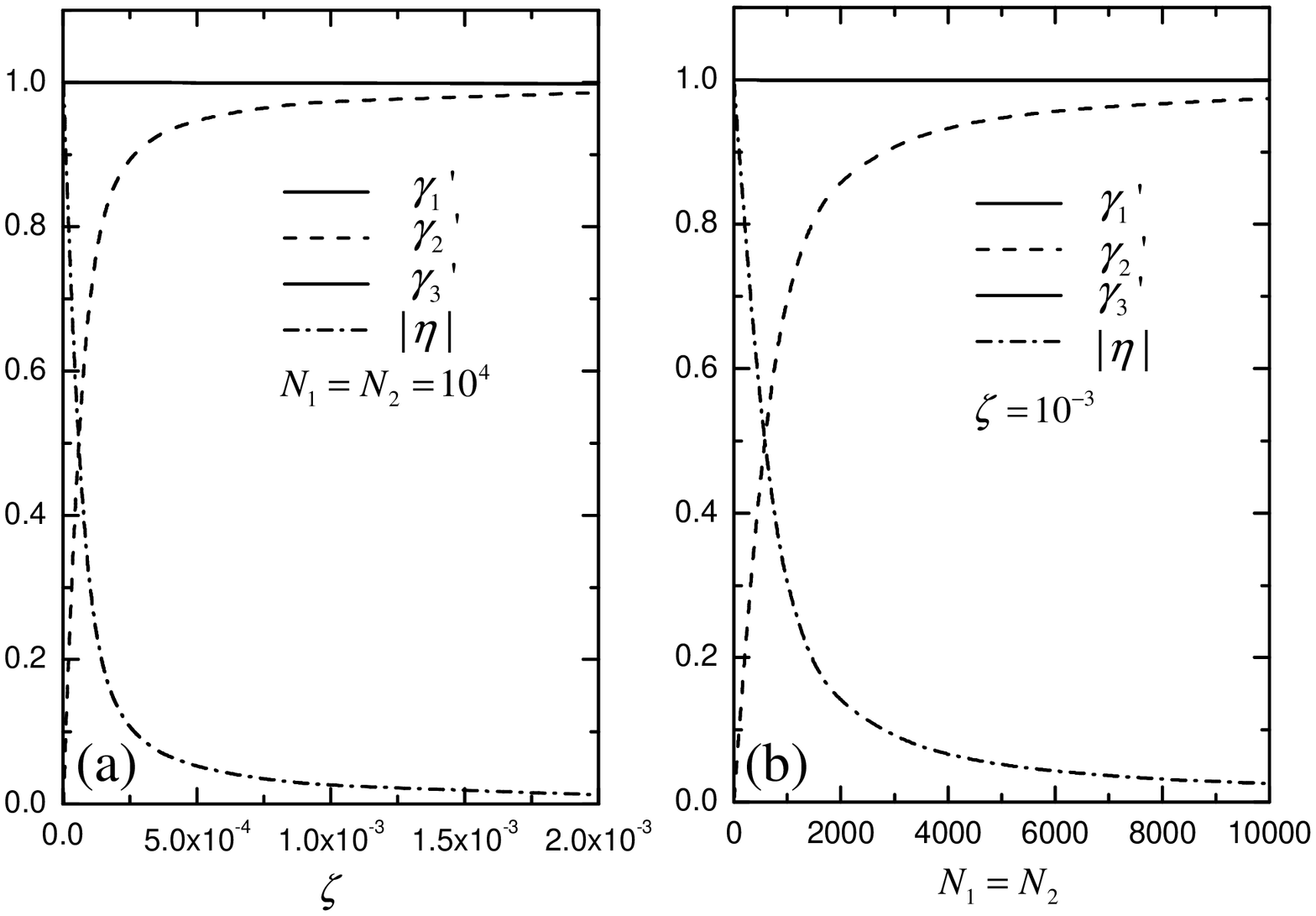}
\caption{Proportion of the non-factorable component in the one-particle
density matrix with different parameters. The relation between $\protect%
\gamma _{1}^{\prime } $, $\protect\gamma _{2}^{\prime }$, $\protect\gamma %
_{3}^{\prime }$, $\left\vert \protect\eta \right\vert $ and $\protect\zeta $
for $N_{1}=N_{2}=10^{4}$ is shown in Fig. 2a, while the relation between $%
\protect\gamma _{1}^{\prime }$, $\protect\gamma _{2}^{\prime }$, $\protect%
\gamma _{3}^{\prime }$, $\left\vert \protect\eta \right\vert $ and $%
N_{1}=N_{2}$ for $\protect\zeta =10^{-3}$ is shown in Fig. 2b. For the case
of $N_{1}=N_{2}$, $\protect\gamma _{1}^{\prime }=\protect\gamma _{3}^{\prime
}$. For $N_{1}\protect\zeta >>1 $, we have $\left\vert \protect\eta %
\right\vert \approx 0$, thus the one-particle density matrix can be
factorized in this situation.}
\end{figure}

For two initially independent and ideal superfluids, because $\zeta \left(
t=0\right) =0$, based on the Schr\H{o}dinger equation, it is easy to verify
that $\zeta \left( t\right) =0$ at any further time. In this situation, $%
\rho \left( \mathbf{x},\mathbf{y},t\right) =\Phi _{1}^{\ast }\left( \mathbf{x%
},t\right) \Phi _{1}\left( \mathbf{y},t\right) +\Phi _{2}^{\ast }\left(
\mathbf{x},t\right) \Phi _{2}\left( \mathbf{y},t\right) $. Obviously, $\rho
\left( \mathbf{x},\mathbf{y},t\right) $ can not be factorized. Thus, the two
superfluids can not be described by a single order parameter even there is
an overlapping between two superfluids.

When the interparticle interaction is taken into account, however, $\zeta
\left( t\right) $ can be a nonzero value. The nonzero value of $\zeta \left(
t\right) $ physically arises from the interparticle interaction of the whole
system. Although generally speaking, $\left\vert \zeta \left( t\right)
\right\vert $ is much smaller than $1$ because $\phi _{1}\phi _{2}^{\ast }$
is an oscillation function about the space coordinate, it is easy to show
based on Eq. (\ref{fin-density-matrix}) that a nonzero value of $\zeta
\left( t\right) $ plays a very important role in the density matrix for
large $N_{1}$ and $N_{2}$. As a general consideration, shown in Fig. 2a is
the relation between $\gamma _{1}^{\prime }$, $\gamma _{2}^{\prime }$, $%
\gamma _{3}^{\prime }$, $\left\vert \eta \right\vert $ and $\zeta $ for $%
N_{1}=N_{2}=10^{4}$, while shown in Fig. 2b is the relation between $\gamma
_{1}^{\prime }$, $\gamma _{2}^{\prime }$, $\gamma _{3}^{\prime }$, $%
\left\vert \eta \right\vert $ and $N_{1}=N_{2}$ for $\zeta =10^{-3}$. We see
that when $N_{1}\left\vert \zeta \right\vert \sim 1$, the factorable
component $\rho _{fac}\left( \mathbf{x},\mathbf{y},t\right) $ gives
significant contribution to $\rho \left( \mathbf{x},\mathbf{y},t\right) $.
In particular, when $N_{1}\left\vert \zeta \right\vert >>1$, one has $%
\left\vert \eta \right\vert \approx 0$, and thus the non-factorable
component $\rho _{non}\left( \mathbf{x},\mathbf{y},t\right) $ can be
omitted. In this situation, the one-particle density matrix can be
factorized, and thus the whole system exhibits the property of ODLRO.
According to the general criterion for Bose superfluid proposed by Penrose
and Onsager, the factorability of the one-particle density matrix means that
two initially independent superfluids have merged into a single superfluid
which can be described by the effective order parameter $\Phi _{e}\left(
\mathbf{x},t\right) $. After two independent superfluids merge into a single
superfluid, we see that the relative phase $\varphi _{c}$ emerges naturally
during the interaction-induced coherence process.

\section{overall energy and \protect\bigskip dynamic evolution}

We now turn to investigate the evolution equations of $\phi _{1}$ and $\phi
_{2}$. The overall energy of the whole system is
\begin{equation}
E=\int dV\left\langle N_{1},N_{2},t\right\vert \left( -\frac{\hbar ^{2}}{2m}%
\widehat{\Psi }^{\dag }\nabla ^{2}\widehat{\Psi }+V_{ext}\widehat{\Psi }%
^{\dag }\widehat{\Psi }+\frac{g}{2}\widehat{\Psi }^{\dag }\widehat{\Psi }%
^{\dag }\widehat{\Psi }\widehat{\Psi }\right) \left\vert
N_{1},N_{2},t\right\rangle ,  \label{overallenergy}
\end{equation}%
where $g=4\pi \hbar ^{2}a/m$ is the coupling constant with $a$ being the
interparticle scattering length. $V_{ext}$ is the external potential of the
system.

After straightforward calculations, the overall energy of the whole system
is
\begin{equation}
E=E_{kin}+E_{pot}+E_{int},
\end{equation}%
where the kinetic energy $E_{kin}$ is given by\bigskip
\begin{eqnarray}
E_{kin} &=&\int \left\langle N_{1},N_{2},t\right\vert \frac{\hbar ^{2}}{2m}%
\nabla \widehat{\Psi }^{\dag }\cdot \nabla \widehat{\Psi }\left\vert
N_{1},N_{2},t\right\rangle dV  \nonumber \\
&=&\int dV\left( \frac{\gamma _{1}\hslash ^{2}}{2m}\nabla \phi _{1}^{\ast
}\cdot \nabla \phi _{1}+\frac{\gamma _{2}\hbar ^{2}}{2m}e^{i\varphi
_{c}}\nabla \phi _{1}^{\ast }\cdot \nabla \phi _{2}\right.  \nonumber \\
&&\left. +\frac{\gamma _{2}\hbar ^{2}}{2m}e^{-i\varphi _{c}}\nabla \phi
_{2}^{\ast }\cdot \nabla \phi _{1}+\frac{\gamma _{3}\hbar ^{2}}{2m}\nabla
\phi _{2}^{\ast }\cdot \nabla \phi _{2}\right) ,
\end{eqnarray}%
while the potential energy is given by%
\begin{eqnarray}
E_{pot} &=&\int \left\langle N_{1},N_{2},t\right\vert V_{ext}\widehat{\Psi }%
^{\dag }\widehat{\Psi }\left\vert N_{1},N_{2},t\right\rangle dV  \nonumber \\
&=&\int V_{ext}\rho \left( \mathbf{r},\mathbf{r},t\right) dV.
\end{eqnarray}%
In addition, the interaction energy $E_{int}$ of the whole system is given by%
\begin{eqnarray}
E_{int} &=&\int \left\langle N_{1},N_{2},t\right\vert \frac{g}{2}\widehat{%
\Psi }^{\dag }\widehat{\Psi }^{\dag }\widehat{\Psi }\widehat{\Psi }%
\left\vert N_{1},N_{2},t\right\rangle dV  \nonumber \\
&=&\frac{g}{2}\int dV\left[ h_{1}\left\vert \phi _{1}\right\vert
^{4}+h_{2}\left\vert \phi _{2}\right\vert ^{4}+h_{3}\left\vert \phi
_{1}\right\vert ^{2}\left\vert \phi _{2}\right\vert ^{2}\right.  \nonumber \\
&&\left. +\mathrm{Re}\left( h_{4}\left\vert \phi _{1}\right\vert ^{2}\phi
_{1}^{\ast }\phi _{2}e^{i\varphi _{c}}+h_{5}\left( \phi _{1}^{\ast }\right)
^{2}\phi _{2}^{2}e^{2i\varphi _{c}}+h_{6}\left\vert \phi _{2}\right\vert
^{2}\phi _{1}^{\ast }\phi _{2}e^{i\varphi _{c}}\right) \right] ,
\label{exact-inter}
\end{eqnarray}%
where the coefficients are%
\begin{eqnarray}
h_{1} &=&\sum\limits_{i=0}^{N_{2}}\frac{C_{n}^{2}N_{2}!\left(
N_{1}+i-2\right) !N_{1}\left( N_{1}-1\right) }{i!i!\left( N_{1}-2\right)
!\left( N_{2}-i\right) !}\left( 1-\left\vert \zeta \right\vert ^{2}\right)
^{N_{2}-i}\left\vert \zeta \right\vert ^{2i},  \nonumber \\
h_{2} &=&\sum\limits_{i=0}^{N_{2}-2}\frac{C_{n}^{2}N_{2}!\left(
N_{1}+i\right) !}{i!i!N_{1}!\left( N_{2}-i-2\right) !}\left( 1-\left\vert
\zeta \right\vert ^{2}\right) ^{N_{2}-i-2}\left\vert \zeta \right\vert ^{2i},
\nonumber \\
h_{3} &=&\sum\limits_{i=0}^{N_{2}-1}\frac{4C_{n}^{2}N_{2}!\left(
N_{1}+i-1\right) !N_{1}}{i!i!\left( N_{1}-1\right) !\left( N_{2}-i-1\right) !%
}\left( 1-\left\vert \zeta \right\vert ^{2}\right) ^{N_{2}-i-1}\left\vert
\zeta \right\vert ^{2i},  \nonumber \\
h_{4} &=&\sum\limits_{i=0}^{N_{2}-1}\frac{4C_{n}^{2}N_{2}!\left(
N_{1}+i-1\right) !N_{1}}{i!\left( i+1\right) !\left( N_{1}-2\right) !\left(
N_{2}-i-1\right) !}\left( 1-\left\vert \zeta \right\vert ^{2}\right)
^{N_{2}-i-1}\left\vert \zeta \right\vert ^{2i+1},  \nonumber \\
h_{5} &=&\sum\limits_{i=0}^{N_{2}-2}\frac{2C_{n}^{2}N_{2}!\left(
N_{1}+i\right) !}{i!\left( i+2\right) !\left( N_{1}-2\right) !\left(
N_{2}-i-2\right) !}\left( 1-\left\vert \zeta \right\vert ^{2}\right)
^{N_{2}-i-2}\left\vert \zeta \right\vert ^{2i+2},  \nonumber \\
h_{6} &=&\sum\limits_{i=0}^{N_{2}-2}\frac{4C_{n}^{2}N_{2}!\left(
N_{1}+i\right) !}{i!\left( i+1\right) !\left( N_{1}-1\right) !\left(
N_{2}-i-2\right) !}\left( 1-\left\vert \zeta \right\vert ^{2}\right)
^{N_{2}-i-2}\left\vert \zeta \right\vert ^{2i+1}.
\end{eqnarray}%
\bigskip

It is well known that the action principle is quite useful to derive the
time-dependent GP equation for a single Bose superfluid. By using the
ordinary action principle and the above overall energy, one can get the
following coupled evolution equations for $\phi _{1}$ and $\phi _{2}$:
\begin{eqnarray}
i\hslash \frac{\partial \phi _{1}}{\partial t} &=&\frac{1}{N_{1}}\frac{%
\delta E}{\delta \phi _{1}^{\ast }},  \nonumber \\
i\hslash \frac{\partial \phi _{2}}{\partial t} &=&\frac{1}{N_{2}}\frac{%
\delta E}{\delta \phi _{2}^{\ast }},  \label{evo-equation}
\end{eqnarray}%
where $\delta E/\delta \phi _{1}^{\ast }$ and $\delta E/\delta \phi
_{2}^{\ast }$ are functional derivatives.

Although the coupled evolution equations (\ref{evo-equation}) are quite
complex, for the case of $N_{1}\left\vert \zeta \right\vert >>1$, $%
N_{2}\left\vert \zeta \right\vert >>1$ and $N_{1}\sim N_{2}$, there is a
very concise approximate evolution equation. When these conditions are
satisfied, the overall energy of the whole system can be approximated very
well as%
\begin{equation}
E_{app}^{\prime }=E_{kin}^{\prime }+E_{pot}^{\prime }+E_{int}^{\prime },
\label{S-app-energy}
\end{equation}%
where%
\begin{equation}
E_{kin}^{\prime }=\frac{\hbar ^{2}}{2m}\int \nabla \Phi _{e}^{\ast }\cdot
\nabla \Phi _{e}dV,  \label{appro-inter}
\end{equation}%
\begin{equation}
E_{pot}^{\prime }=\int V_{ext}\left\vert \Phi _{e}\right\vert ^{2}dV,
\end{equation}%
\begin{equation}
E_{int}^{\prime }=\frac{g}{2}\int dV\left\vert \Phi _{e}\right\vert ^{4}.
\label{eappr-inter}
\end{equation}

For the cases of $N_{1}\left\vert \zeta \right\vert >>1$, $N_{2}\left\vert
\zeta \right\vert >>1$ and $N_{1}\sim N_{2}$, it is easy to verify that $%
E_{kin}^{\prime }\approx E_{kin}$ and $E_{pot}^{\prime }\approx E_{pot}$
based on the analogous analyses about the effective order parameter that $%
\rho \left( \mathbf{x},\mathbf{y},t\right) \approx \Phi _{e}^{\ast }\left(
\mathbf{x},t\right) \Phi _{e}\left( \mathbf{y},t\right) $ in this situation.
For $N_{1}\left\vert \zeta \right\vert >>1$, $N_{2}\left\vert \zeta
\right\vert >>1$ and $N_{1}\sim N_{2}$, one can also prove the result of $%
E_{int}^{\prime }\approx E_{int}$. Based on Eq. (\ref{eappr-inter}), $%
E_{int}^{\prime }$ can be expanded as:
\begin{eqnarray}
E_{int}^{\prime } &=&\frac{g}{2}\int dV\left[ \beta _{1}\left\vert \phi
_{1}\right\vert ^{4}+\beta _{2}\left\vert \phi _{2}\right\vert ^{4}+\beta
_{3}\left\vert \phi _{1}\right\vert ^{2}\left\vert \phi _{2}\right\vert
^{2}\right.   \nonumber \\
&&\left. +\mathrm{Re}\left( \beta _{4}\left\vert \phi _{1}\right\vert
^{2}\phi _{1}^{\ast }\phi _{2}e^{i\varphi _{c}}+\beta _{5}\left( \phi
_{1}^{\ast }\right) ^{2}\phi _{2}^{2}e^{2i\varphi _{c}}+\beta _{6}\left\vert
\phi _{2}\right\vert ^{2}\phi _{1}^{\ast }\phi _{2}e^{i\varphi _{c}}\right) %
\right] ,
\end{eqnarray}%
where
\begin{eqnarray}
\beta _{1} &=&\gamma _{1}^{2},  \nonumber \\
\beta _{2} &=&\gamma _{3}^{2},  \nonumber \\
\beta _{3} &=&4\gamma _{1}\gamma _{3},  \nonumber \\
\beta _{4} &=&4\gamma _{1}\sqrt{\gamma _{1}\gamma _{3}},  \nonumber \\
\beta _{5} &=&2\gamma _{1}\gamma _{3},  \nonumber \\
\beta _{6} &=&4\gamma _{3}\sqrt{\gamma _{1}\gamma _{3}}.
\end{eqnarray}%
To compare with the exact expression of the interaction energy given by Eq. (%
\ref{exact-inter}), Fig. 3 shows the relation between $h_{i}/\beta _{i}$ ($%
i=1,\cdots ,6$) and $\zeta $ for $N_{1}=N_{2}=10^{3}$. It is shown clearly
that for $N_{1}\left\vert \zeta \right\vert >>1$ and $N_{2}\left\vert \zeta
\right\vert >>1$, $h_{i}/\beta _{i}\approx 1$, and thus $E_{int}^{\prime
}\approx E_{int}$.
\begin{figure}[tbp]
\includegraphics[width=0.5\linewidth,angle=270]{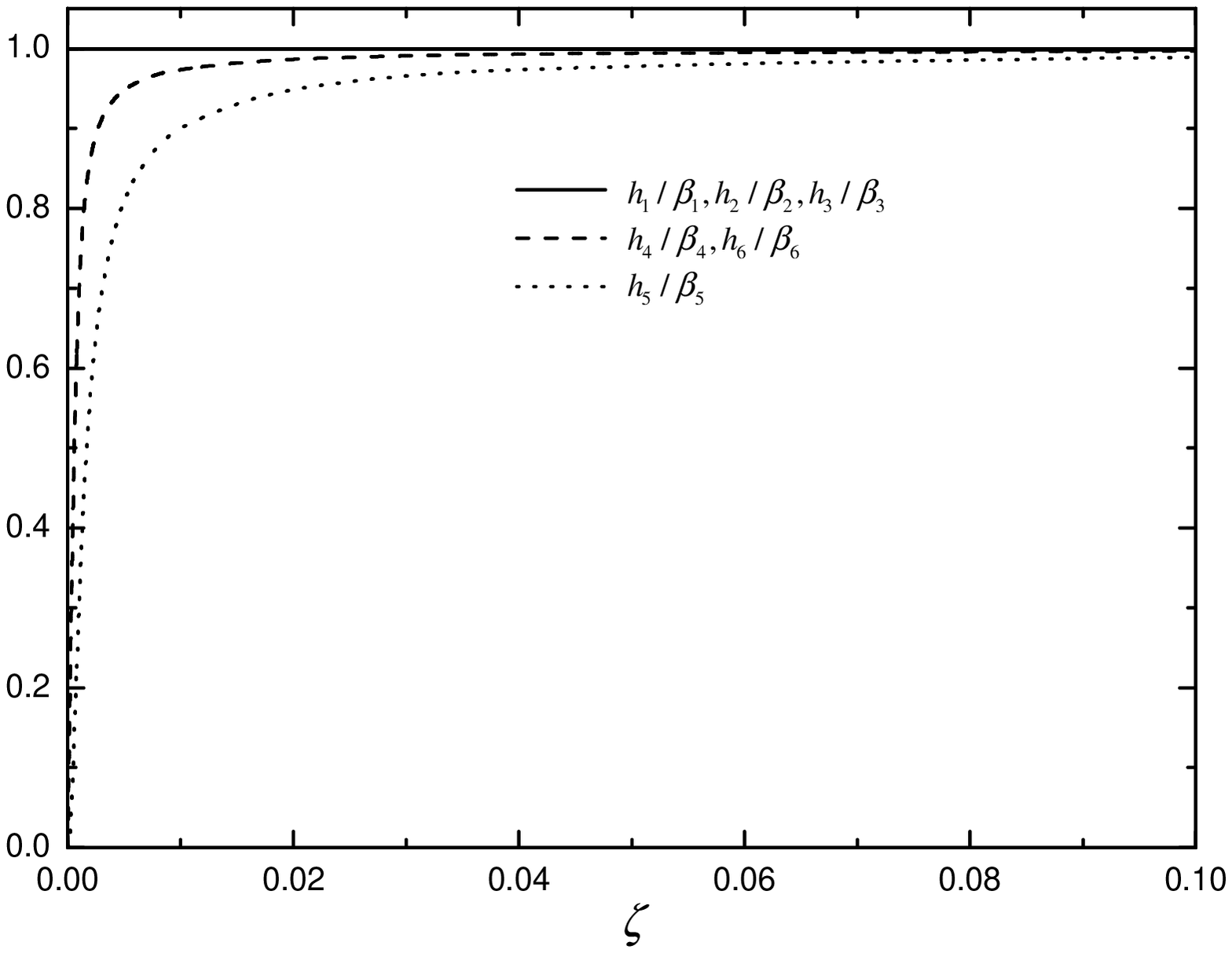}
\caption{Shown is the relation between $h_{i}/\protect\beta _{i}$ ($%
i=1,\cdots ,6$) and $\protect\zeta $ for $N_{1}=N_{2}=10^{3}$. We see that
for $N_{1}\left\vert \protect\zeta \right\vert >>1$ and $N_{2}\left\vert
\protect\zeta \right\vert >>1$, $h_{i}/\protect\beta _{i}\approx 1$, which
means that $E_{int}^{\prime }\approx E_{int}$.}
\end{figure}

Therefore, the overall energy can be approximated well as

\begin{equation}
E\simeq -\frac{\hbar ^{2}}{2m}\int \Phi _{e}^{\ast }\nabla ^{2}\Phi
_{e}dV+\int V_{ext}\left\vert \Phi _{e}\right\vert ^{2}dV+\frac{g}{2}\int
dV\left\vert \Phi _{e}\right\vert ^{4},  \label{app-energy}
\end{equation}%
where $\Phi _{e}\left( \mathbf{x},t\right) $ is the effective order
parameter given by Eq. (\ref{effective-order}). Based on this approximate
energy expression, it is easy to get the following evolution equation for $%
\Phi _{e}\left( \mathbf{x},t\right) $:
\begin{equation}
i\hslash \frac{\partial \Phi _{e}}{\partial t}\simeq -\frac{\hslash ^{2}}{2m}%
\nabla ^{2}\Phi _{e}+V_{ext}\Phi _{e}+g\left\vert \Phi _{e}\right\vert
^{2}\Phi _{e}.  \label{app-equation}
\end{equation}%
It is quite interesting to note that the above equation has the same form as
the standard GP equation \cite{GP}.

Because there is no exact analytic solution for Eq. (\ref{evo-equation}),
here we give the solution based on numerical calculations. We consider the
one-dimensional dynamic process when two initially independent condensates
in dilute atomic gases are weakly connected through adiabatically decreasing
the height of the barrier separating two condensates. The time-dependent
double-well potential is assumed as
\begin{equation}
V_{ext}\left( x,t\right) =\frac{1}{2}m\omega _{x}^{2}x^{2}+U\left( t\right)
e^{-x^{2}/\Delta _{x}^{2}},
\end{equation}%
where $U\left( t\right) =U_{0}e^{-\theta t}+U_{1}$. The first term is the
external potential due to a magnetic trap or an optical trap, while the
second term is the central barrier due to a laser beam separating two
condensates. When the central barrier separating two condensates is
sufficiently high so that there is no tunneling current, one may prepare two
completely independent (rather than coherently separated) condensates by
directly cooling the dilute gases in the double-well trap. In the last ten
years, the remarkable experimental advances \cite{BEC} on Bose condensate in
dilute gases make it be quite promising to experimentally test the
theoretical predication in this work.

In the numerical calculations, we introduce the dimensionless parameters $%
x_{0}=x/l_{x}$, $\tau =E_{l}t/\hslash $, $g^{\prime }=g/E_{l}l_{x}$. Here $%
l_{x}=\sqrt{\hslash /m\omega _{x}}$ and $E_{l}=\hbar ^{2}/2ml_{x}^{2}$. The
particle number is $N_{1}=2\times 10^{4}$ and $N_{2}=10^{4}$. In addition, $%
U_{0}=400E_{l}$, $U_{1}=20E_{l}$, $\theta =2E_{l}/\hbar $, $\Delta
_{x}=l_{x} $ and $g^{\prime }N_{1}=50$. For these parameters, the tunneling
effect can be omitted for two independent condensates at the initial time.
When the central barrier due to the laser beam is decreased, there is
particle current between two condensates.

In the numerical calculations, first we get the ground state wave functions $%
\phi _{1}$ and $\phi _{2}$ at the initial time. Then, by numerically
calculating the coupled equations (\ref{evo-equation}), we obtain the
evolution of $\zeta \left( t\right) $ based on the numerical results of $%
\phi _{1}$ and $\phi _{2}$. From $\zeta \left( t\right) $, we give the
evolution of $\left\vert \eta \right\vert $ in Fig.4. We have also confirmed
in the numerical calculations that, for $g^{\prime }=0$, the numerical
result of $\left\vert \zeta \right\vert $ can be regarded as zero because it
is smaller than $10^{-10}$. This shows clearly that the nonzero value of $%
\left\vert \zeta \right\vert $ physically arises from the interatomic
interaction, rather than the error in the numerical calculations. In fact,
if we assume that $\zeta \left( t\right) $ is always zero with the
development of time, it is easy to show that this is an inconsistent
assumption because with this assumption we prove both in the numerical
calculations and analytic analysis that $\left\vert \zeta \right\vert $ will
increase from zero after the overlapping between two condensates. In the
inset of Fig.4, we give the numerical result of $N_{1}\left( t\right) /N$
which shows clearly the Josephson effect. Our numerical results show that $%
\left( N_{1}\left( t\right) +N_{2}\left( t\right) \right) /N$ is always
equal to $1$ with an error below $10^{-11}$ which confirms further our
numerical calculations.

\begin{figure}[tbp]
\includegraphics[width=0.6\linewidth,angle=270]{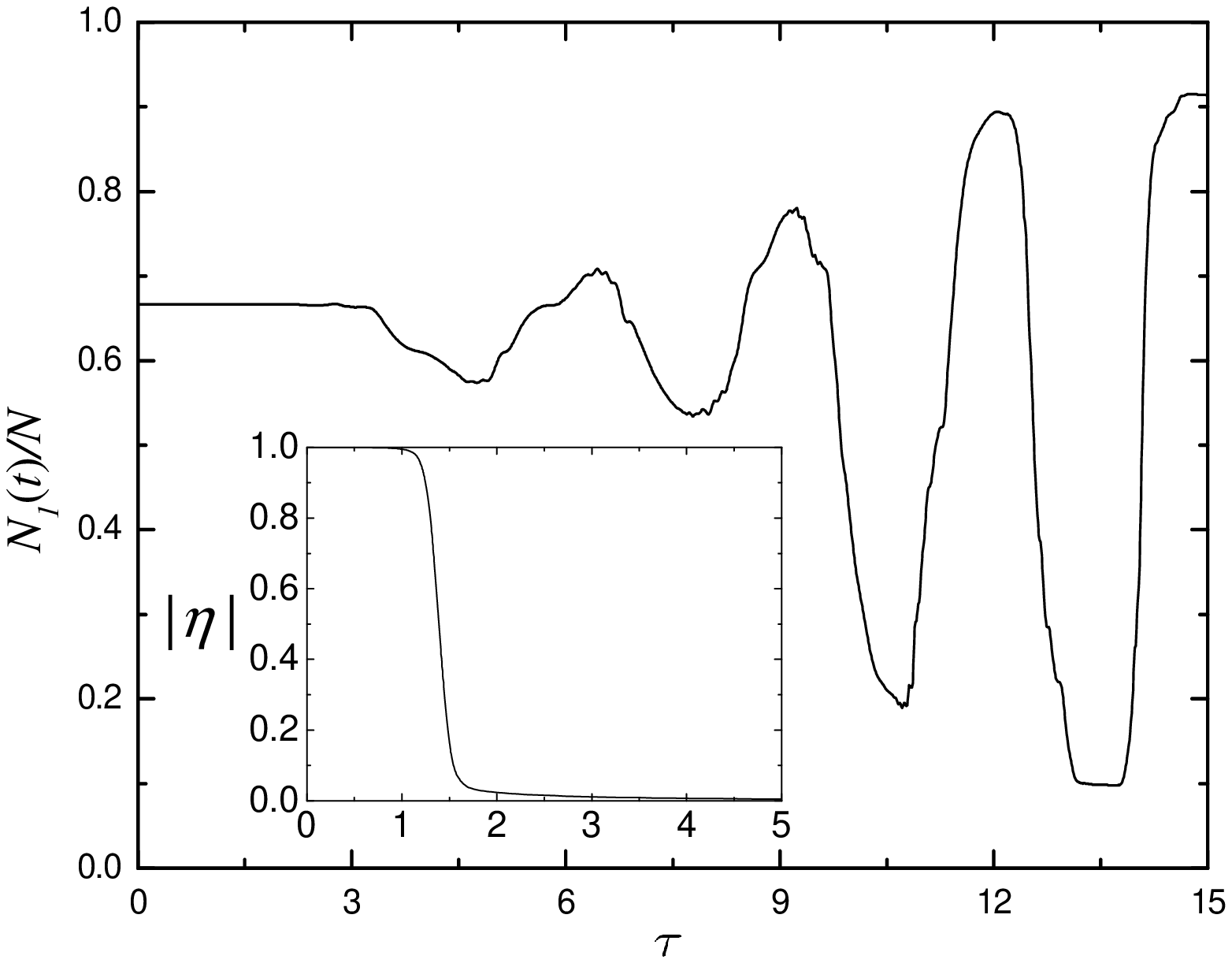}
\caption{Quantum merging process shown through the evolution of $\left\vert
\protect\eta \right\vert $ and the Josephson effect. The evolution of $%
N_{1}/N$ shows clearly the Josephson effect. Due to the particle current
between two initially independent condensates, as shown in the inset, $%
\left\vert \protect\eta \right\vert $ decreases significantly with the
development of time. At $\protect\tau >3$, $\left\vert \protect\eta %
\right\vert \approx 0 $, thus two initially independent condensates have
merged into a single condensate.}
\end{figure}

Although the Josephson effect shown in Fig.4 is the numerical results of Eq.
(\ref{evo-equation}), one can understand easily this effect with the
effective order parameter. Due to the quantum merging process, when the
central barrier separating two independent condensates is decreased, two
initially independent condensates will merge into a single condensate
described by the effective order parameter $\Phi _{e}$ whose evolution is
determined by GP equation (\ref{app-equation}). Thus, it is natural that
there is coherent Josephson current for two initially independent
interacting condensates.

\section{\protect\bigskip summary and discussion}

In summary, we investigate the dynamic process of the whole system when the
barrier separating two initially independent Bose superfluids is
adiabatically decreased so that the two initially separated and independent
superfluids are weakly connected. When the interparticle interaction is
considered, we show that there is an effective order parameter for the whole
system under appropriate condition. Compared with previous interesting
studies \cite{Zapata,Yi,Mebrahtu} about quantum merging, in the present
work, we consider this problem by stressing the non-orthogonal property
between the wave functions for different condensates after their overlapping
and in the presence of interparticle interaction. In particular, it is found
that the effective order parameter satisfies the ordinary Gross-Pitaevskii
equation, which means that there is also Josephson effect for two initially
independent Bose superfluids. This result for the effective order parameter
makes our theory can be tested and applied directly in future experiments
about quantum merging process, such as the experiment about Josephson effect
for independent Bose superfluids.

We stress here again that, in our theory, the quantum depletion originating
from the elementary excitations at zero temperature is omitted in the
initial state (\ref{initial-state}). Based on the Bogoliubov theory of the
elementary excitations \cite{Pethick}, the number of particles due to the
elementary excitations is of the order of $\left( a/\overline{l}\right)
^{3/2}$ and thus the quantum depletion is negligible for Bose condensate in
dilute gases. There is another reason why the elementary excitations can be
omitted in the effective order parameter. A simple analysis shows that $%
\left\langle \phi _{1}|\phi _{k}\right\rangle $ and $\left\langle \phi
_{2}|\phi _{k}\right\rangle $ (here $\phi _{k}$ is the normalized wave
function of the elementary excitations) are of the order of $\left\vert
\zeta \right\vert e^{-(k-2\pi /L)^{2}}$ with $k$ and $L$ being respectively
the wave number of the elementary excitations and spatial size of the Bose
superfluid. This exponential decay of $\left\langle \phi _{1}|\phi
_{k}\right\rangle $ and $\left\langle \phi _{2}|\phi _{k}\right\rangle $
originates from the integral where there is spatially oscillating phase
factor in the wave functions of the elementary excitations and Bose
superfluid. Thus, even there are elementary excitations, its contribution to
the effective order parameter is negligible. Although the present theory can
not give quantitative predication for liquid superfluid of $^{4}\mathrm{He}$
because the quantum depletion is very important, we believe that the quantum
merging process means that after two separate tanks are connected by a pipe,
two superfluids of $^{4}\mathrm{He}$ can merge into a single superfluid.

For the quantum state given by Eq. (1), with the development of time, the
wave functions $\phi _{1}$ and $\phi _{2}$ are no more orthogonal in the
presence of interparticle interaction. If we use the orthogonal bases $\phi
_{1}$ and $\phi _{2}^{\prime }$, the quantum state of the whole system is $%
\left\vert N_{1},N_{2}\right\rangle \sim (\widehat{a}_{1}^{\dag })^{N_{1}}(%
\widehat{k}/\beta ^{\ast }+\zeta \widehat{a}_{1})^{N_{2}}\left\vert
0\right\rangle $. We see that the number of particles in the orthogonal
modes $\phi _{1}$ and $\phi _{2}^{\prime }$ are no more definite. This
quantum state becomes a superposition of different number of particles in
the orthogonal modes $\phi _{1}$ and $\phi _{2}^{\prime }$. This implies
strongly that in some sense our theory is equivalent to the Gutzwiller type
approach \cite{Gutzwiller} where the coupling between different modes leads
to the coherent transfer of particles between different orthogonal modes.

It is straightforward to generalize the present work to the quantum merging
process of several independent Bose superfluids. It is possible that this
quantum merging process contributes to our understanding of the Bose
condensation process. During the evaporative cooling process, firstly there
would be a series of independent subcondensates formed from the thermal
cloud. Due to the quantum merging process, these independent subcondensates
will overlap and finally merge into a single condensate with well spatial
coherence property. During the evaporative cooling process, due to the
thermal equilibrium of the whole system, the thermal atoms continuously jump
into the ground state. The quantum merging process has the effect that the
atoms in the ground sate merge into a single condensate which has well
spatial and phase coherence property.

\begin{acknowledgments}
This work is supported by NSFC under Grant Nos. 10474117, 10474119 and NBRPC
under Grant Nos. 2005CB724508, 2001CB309309, and also funds from Chinese
Academy of Sciences.

Electronic address: xionghongwei@wipm.ac.cn
\end{acknowledgments}

\end{document}